\documentclass[fleqn,usenatbib]{mnras}

\usepackage{newtxtext,newtxmath}

\usepackage[T1]{fontenc}
\usepackage{xspace}

\DeclareRobustCommand{\VAN}[3]{#2}
\let\VANthebibliography\thebibliography
\def\thebibliography{\DeclareRobustCommand{\VAN}[3]{##3}\VANthebibliography}

\usepackage{graphicx}	
\usepackage{amsmath}	
\usepackage{xcolor}

\newcommand{\ocen}{$\omega$ Cen\xspace}
\newcommand{\Omegab}{\Omega_\mathrm{b}}

\title[Bar-induced migration of $\omega$ Cen]{Bar-induced migration of $\omega$ Centauri away from Gaia Sausage-Enceladus}

\author[A. M. Dillamore et al.]{
Adam M. Dillamore,$^{1}$\thanks{E-mail: a.dillamore@ucl.ac.uk (AMD)}
Hanyuan Zhang$^{2}$
and Vasily Belokurov$^{2}$
\\
$^{1}$Department of Physics and Astronomy, University College London, London, WC1E 6BT, UK\\
$^{2}$Institute of Astronomy, University of Cambridge, Madingley Road, Cambridge CB3 0HA, UK
}

\date{Accepted XXX. Received YYY; in original form ZZZ}

\pubyear{\the\year{}}

\begin{document}
\label{firstpage}
\pagerange{\pageref{firstpage}--\pageref{lastpage}}
\maketitle

\begin{abstract}
The globular cluster \ocen has been suggested to have originated in the Gaia Sausage-Enceladus (GSE) merger event, possibly as its nuclear star cluster. However, the present-day orbits of \ocen and the GSE debris are very different. We investigate the scenario in which \ocen originated in the GSE and migrated to its current position due to perturbations from the Galactic bar. The [$\alpha$/M] distributions of stars located between the GSE debris and \ocen in $(L_z,E)$ space tentatively support this scenario, but are not conclusive. We run simulations of the GSE debris and \ocen in a realistic Milky Way potential with a decelerating bar at various present-day pattern speeds. We find that \ocen can indeed be traced back to the phase space region occupied by the GSE debris. However, this would likely require a pattern speed of $\Omegab\lesssim26$~km\,s$^{-1}$\,kpc$^{-1}$, which is much lower than most recent estimates. We conclude that a GSE origin for \ocen is dynamically and chemically plausible, but only with a re-evaluation of the current consensus on the bar's pattern speed.
\end{abstract}




\section{Introduction}
The inner stellar halo of the Milky Way is dominated by debris from a massive early accretion event, commonly known as Gaia-Sausage-Enceladus (GSE). This structure was first identified as a population of relatively metal-rich halo stars on highly radial orbits, producing the characteristic double-lobed, or ``sausage-like'', velocity distribution in the Solar neighbourhood \citep{belokurov2018,helmi2018}. Subsequent work has shown that GSE contributes a large fraction of the nearby accreted halo, is chemically distinct from the in-situ Milky Way populations, and was likely one of the last major mergers experienced by the Galaxy \citep[e.g.][]{mackereth2019,naidu2020,deason2024}. Its progenitor was also massive enough to bring in its own system of globular clusters \citep{MySausageGC,kruijssen2019,Mas19}. The combination of field-star debris and associated clusters makes GSE a natural reference point for interpreting unusual halo clusters whose orbits, chemistry, or stellar populations suggest an accreted origin.

Because GSE brought in a globular-cluster system, it is natural to ask whether one surviving member was the progenitor's own nucleus. A nucleated galaxy contains a compact central star cluster, or nuclear star cluster, at the bottom of its potential well. Such a central star cluster is not simply an ordinary globular cluster projected onto the galaxy centre: it can grow through repeated star formation, gas inflow, and the inspiral and merging of other clusters, and can therefore acquire a broader chemical and age structure than a typical old globular cluster \citep{neumayer2020}. If the Milky Way accretes a nucleated dwarf galaxy, tidal stripping can remove most of the surrounding galaxy while leaving the dense nucleus intact. The remnant nucleus would then appear today as a massive halo cluster, but with stellar populations and chemical enrichment histories inherited from its former position at the centre of a dwarf galaxy. The Sagittarius system and M54 provide the clearest nearby example of this process \citep[e.g.][]{ibata1994,bellazzini2008}. Other massive and chemically complex Milky Way clusters have also been proposed as stripped nuclear star clusters, including \ocen, NGC 6273, and possibly NGC 6934 \citep{pfeffer2021}.

The strongest candidate for such a remnant in the Milky Way halo is \ocen. Although traditionally classified as the most massive Galactic globular cluster, \ocen is exceptional in several respects: it has a broad and multi-peaked metallicity distribution, multiple stellar populations, internal abundance variations in iron-peak and neutron-capture elements, and evidence for a complex age-metallicity structure \citep[e.g.][]{norris2004,johnson2010,gratton2011,villanova2014,Nitschai2024,Alvarez2024}. These properties have long motivated the idea that \ocen is the stripped nucleus of an accreted dwarf galaxy rather than a normal globular cluster \citep[e.g.][]{hilker2000,bekki2003,akbaba2026}. The identity of the parent galaxy, however, remains uncertain. Its retrograde orbit motivated links to retrograde halo substructure, including the Sequoia event \citep{myeong2019_sequoia}. More recent chemo-dynamical studies have instead argued that GSE is a plausible, and perhaps preferred, parent system because the metallicity scale, inferred progenitor mass, and broader accretion context are compatible with a nuclear remnant from the GSE galaxy \citep{pfeffer2021,limberg2022,laporteorkney2026,souza2026}. The main difficulty with this association is dynamical: \ocen now lies at lower orbital energy and more retrograde angular momentum than the bulk of the observed GSE debris.

This phase-space offset is hard to produce in an axisymmetric Galactic potential, even with dynamical friction \citep{moreno2022}. However, it has become clear in recent years that the Galactic bar causes significant perturbations to orbits in the inner stellar halo \citep{dillamore2023,dillamore2024b,dillamore2025,dillamore2025b,tomlinson2026,deleo2026,gherghinescu2026}. The bar's pattern speed $\Omegab$ (rotation rate) is likely to be decelerating \citep{chiba2021,zhang25}, a process which can cause migration of globular clusters via resonances \citep{dillamore2024b}. These include the retrograde 1:1 resonance, where the orbital frequencies of a particle satisfy $\Omega_\phi-\Omegab+\Omega_r=0$. \citet{tomlinson2026} showed that this resonance can cause significant scattering of orbits along its length in $(L_z,E)$ space \citep[on lines of gradient $\mathrm{d}E/\mathrm{d}L_z=\Omegab$; see][]{dillamore2025b}. This process could be responsible for deflecting \ocen onto lower energy, more retrograde orbits \citep{laporteorkney2026}. In this \textit{Letter} we investigate this scenario by running simulations in a realistic Milky Way potential with a decelerating bar. We determine the range of pattern speeds compatible with the migration of \ocen from the GSE debris, and thus assess how plausible this scenario is. We also analyse the chemical composition of retrograde substructure between the GSE debris and \ocen, extending the analysis of \citet{laporteorkney2026}.

The rest of this \textit{Letter} is arranged as follows. In Section~\ref{section:data} we describe the data used, and analyse the chemistry of retrograde substructure. Section~\ref{section:simulations} discusses our simulations and the constraint on $\Omegab$. Finally we discuss and summarise our conclusions in Section~\ref{section:conclusions}.

\section{Data}\label{section:data}
\noindent\textbf{\ocen coordinates.} We take the proper motion of \ocen from \textit{Gaia} EDR3 measurements \cite{gaia_edr3} published by \citet{vasiliev2021}. We use the distance and line-of-sight velocity estimates from \citet{baumgardt2021} and \citep{baumgardt2019} respectively, and on-sky position from \citet{harris2010}. We assume the uncertainties in distance and each velocity component are Gaussian-distributed with standard deviations equal to the errors reported by \citet{vasiliev2021}. We represent the present-day phase space position of \ocen by drawing $N=10^3$ samples from these distributions. We transform into Galactocentric coordinates using the default position and velocity of the Sun in \textsc{astropy} \citep{astropy:2013,astropy:2018}, specifically position $\boldsymbol{x}_\odot=(-8.122,0,0.0208)\,$kpc and velocity $\boldsymbol{v}_\odot=(12.9, 245.6, 7.78)\,$km\,s$^{-1}$ \citep{schonrich2010,gravity2018,bennett2019}. We also assume that the Sun is located at an angle of $28^\circ$ to the Galactic bar \citep{bland-hawthorn2016,hunter2024}. Henceforth we work in a left-handed coordinate system such that the disc and \ocen have positive and negative $L_z$ respectively. In this coordinate system we refer to the Cartesian coordinates of the \ocen location samples as $(\boldsymbol{x}_i,\boldsymbol{v}_i)$.

\noindent\textbf{Chemistry of retrograde substructure.} Section~4.2 of \citet{laporteorkney2026} argues that the chemical structure of the retrograde halo in $(L_z,E)$ may preserve a link between the main Gaia-Sausage-Enceladus (GSE) debris cloud and \ocen. Their fig.~5 shows that the core of the GSE is more metal-rich than its surroundings and highlights a diagonal metal-poor corridor extending from the GSE cloud towards lower energy and more retrograde angular momentum, in the general direction of \ocen. Motivated by that result, we revisit the same region of integrals-of-motion space and ask whether the putative corridor also carries a distinctive $\alpha$-abundance signature.

\begin{figure*}
  \centering
  \includegraphics[width=\textwidth]{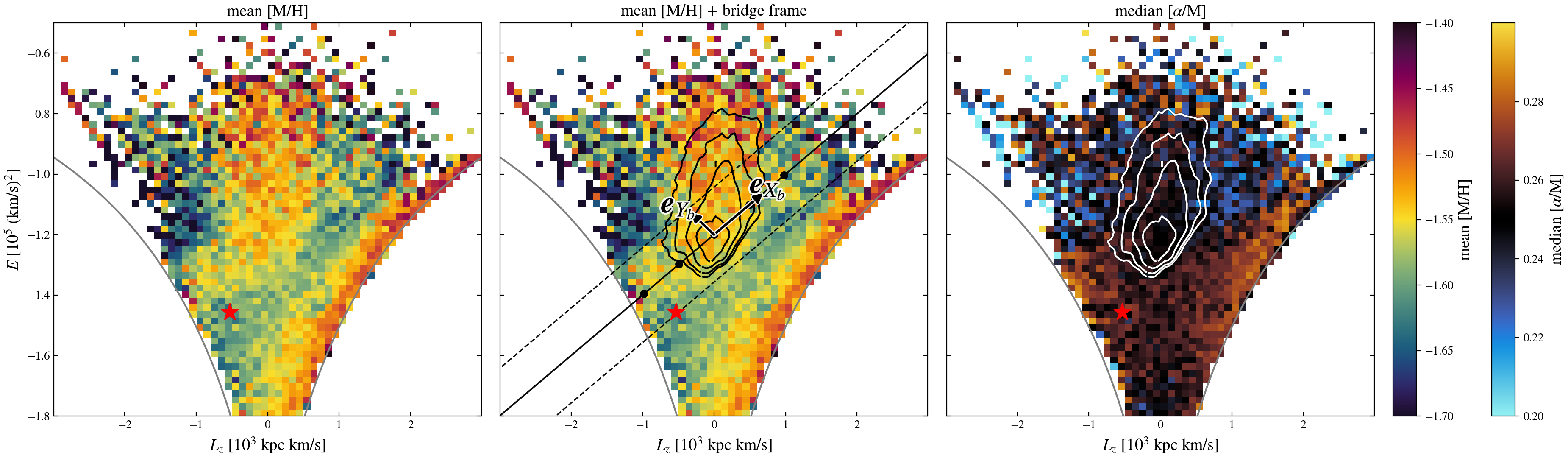}
  \caption{Chemical maps in the $(L_z,E)$ plane, following fig.~5 of \citet{laporteorkney2026} with Gaia XP abundances \citep{li2024aspgap}. Left: mean $[M/H]$, showing the metal-rich GSE cloud and the diagonal corridor towards lower energy and more retrograde $L_z$. Middle: the same map with GSE contours from \citet{belokurov_chevrons}, the present-day \ocen location (red star), and the bridge coordinates used in Fig.~\ref{fig:bridge_alpha}. \textbf{The bridge-aligned coordinate frame is centred on $(L_z,E)=(0,-1.2)$; arrows show the $X_b$ and $Y_b$ basis directions, with $\boldsymbol{e}_{X_b}\parallel(\Delta L_z,\Delta E)=(1.0,0.2)$.} Black lines mark $Y_b=-0.4,0.0,+0.4$, and black points mark $X_b=\{-1.0,-0.5,0.0,0.5,1.0\}$ at $Y_b=0$. Right: median $[\alpha/\rm{M}]$ in the same bins; the GSE cloud is relatively low-$\alpha$, while the corridor is slightly more $\alpha$-enhanced.}
  \label{fig:metallicity}
\end{figure*}

Figure~\ref{fig:metallicity} was constructed to mimic the chemical map in \citet{laporteorkney2026}. For the chemistry, we use the Gaia XP abundance catalogue of \citet{li2024aspgap}, adopting the published $[\rm{M/H}]$ and $[\alpha/\rm{M}]$ labels stored in the {\tt moh\_xp} and {\tt aom\_xp} columns. We use this catalogue because its homogeneous all-sky coverage and very large sample size make it possible to map mean metallicity and median $\alpha$ trends across the $(L_z,E)$ plane with adequate number statistics, even if the abundance precision for individual stars is lower than in smaller spectroscopic surveys. The parent sample is restricted to stars with $\varpi/\sigma_\varpi>5$, ${\tt phot\_bp\_mean\_mag}<16.5$, ${\tt teff\_xp}<5200$, ${\tt logg\_xp}<3.5$, and heliocentric distance $r<1000$ kpc. We compute $(E,L_z)$ in the axisymmetric \citet{hunter2024} potential and then apply the additional cuts $-1.9<[\rm{M/H}]<-1.3$ and $[\alpha/\rm{M}]>0.1$ in order to follow the bridge-band selection used by \citet{laporteorkney2026}. The left-hand panel shows the mean metallicity in $60\times60$ bins across the fixed $(L_z,E)$ window, with bins containing fewer than three stars blanked. The middle panel overlays the GSE contours from \citet{belokurov_chevrons}, the present-day location of \ocen, and the corridor-aligned coordinate system introduced below. The right-hand panel complements the metallicity map with the median $[\alpha/\rm{M}]$ measured in the same bins. In this panel the GSE cloud appears as a relatively low-$\alpha$ feature compared to much of the surrounding halo, whereas the corridor identified by \citet{laporteorkney2026} is, if anything, slightly more $\alpha$-enhanced than the core of the GSE itself. This already suggests that any chemical bridge between the GSE debris and \ocen is not simply a smooth continuation of the low-$\alpha$ GSE sequence.

To examine the corridor more directly, we project the $(L_z,E)$ plane into a bridge frame. We place the origin at $(L_z,E)=(0,-1.2)$ and define the positive $X_b$ direction by the basis vector $\boldsymbol{e}_{X_b}\parallel(\Delta L_z,\Delta E)=(1.0,0.2)$, which follows the slope of the candidate bridge in Fig.~\ref{fig:metallicity}. The $Y_b$ direction is perpendicular to $X_b$ in the displayed metric of Fig.~\ref{fig:metallicity}. This frame lets us slice the halo parallel and perpendicular to the proposed corridor and track how the $[\alpha/\rm{M}]$ distribution changes with metallicity.

\begin{figure}
  \centering
  \includegraphics[width=\columnwidth]{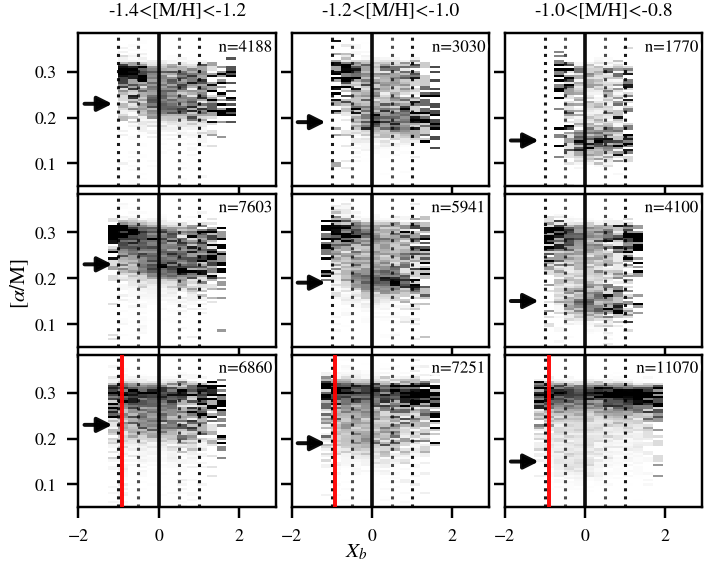}
  \caption{Column-normalized density in the $(X_b,[\alpha/\rm{M}])$ plane for the bridge coordinate system, defined in the text and illustrated in Fig.~\ref{fig:metallicity}. Rows correspond, from top to bottom, to corridors centred on $Y_b=+0.4$, $0.0$, and $-0.4$, each with half-width $0.2$; columns correspond to $0.2$-dex metallicity bins. The solid vertical line marks $X_b=0$, dotted lines mark $X_b=\pm0.5$ and $\pm1.0$, arrows mark the low-$\alpha$ GS/E sequence, and the red line in the bottom row marks the \ocen value $X_b=-0.906$. The low-$\alpha$ sequence is clearest near the GS/E cloud and only tentatively extends towards \ocen.}
  \label{fig:bridge_alpha}
\end{figure}

Figure~\ref{fig:bridge_alpha} shows the resulting $[\alpha/\rm{M}]$ distributions as a function of $X_b$. Rows correspond to three parallel corridors centred on $Y_b=+0.4$, $0.0$, and $-0.4$, each with half-width $0.2$, while columns use three coarse metallicity bins: $-1.4<[\rm{M/H}]<-1.2$, $-1.2<[\rm{M/H}]<-1.0$, and $-1.0<[\rm{M/H}]<-0.8$. Each panel is displayed as a column-normalized density map in the $(X_b,[\alpha/\rm{M}])$ plane, so that narrow sequences remain visible even where the total number of stars varies strongly with $X_b$. In most panels the GSE debris is conspicuous near $X_b\approx0$ as a low-$\alpha$ component. This sequence shifts to slightly lower $[\alpha/\rm{M}]$ towards the more metal-rich columns on the right, as expected from earlier studies of GSE chemistry \citep{helmi2018,mackereth2019}; the horizontal arrows mark this progression. If one instead scans from top to bottom across the rows, the GSE sequence is strongest in the upper rows and becomes fainter in the bottom row, consistent with earlier GSE studies and with the density map of \citet{belokurov_chevrons}. 

In this frame \ocen has $(X_b,Y_b)=(-0.906,-0.383)$, placing it in the corridor centred on $Y_b=-0.4$, so the bottom row alone carries a vertical red line at $X_b=-0.906$. The low-$\alpha$ sequence is weaker in this row than in the upper two, but the more metal-poor columns show at least tentative hints of low-$\alpha$ stars near the location of \ocen marked by the red line. Most panels nevertheless do not show an obvious continuation of the same low-$\alpha$ sequence to $X_b\lesssim-0.5$, where one might hope to trace a direct extension towards \ocen. Taken together, the maps do not support a simple, chemically uniform bridge across all corridor slices, but neither do they rule out a weaker continuation directly towards \ocen. The present chemical evidence is therefore suggestive, but not yet decisive, in linking the GSE cloud to \ocen through the proposed corridor.

\section{Simulations}\label{section:simulations}
We now run simulations using \textsc{agama} \citep{agama} of a GSE-like population and \ocen itself in a Milky Way potential with a decelerating bar. This will help to determine whether the bar is capable of causing migration of \ocen from the GSE debris to its current orbit.

\noindent\textbf{Slowing barred potential.} We integrate orbits in a modified version of the Milky Way barred potential from \citet{hunter2024}. We consider decelerating bars which evolve according to the prescription used by \citet{dillamore2024,dillamore2024b,dillamore2025}. The bar increases in amplitude between times $t=0$ and $t_1\approx1$\,Gyr, then begins to smoothly decelerate until $t_2\approx2$\,Gyr. Between $t=t_2$ and the end of the simulation at $t_\mathrm{f}\approx8$\,Gyr, the bar's dimensionless deceleration rate $\eta\equiv-\dot{\Omega}_\mathrm{b}/\Omegab^2$ is held constant at $\eta=0.003$, consistent with \citet{chiba2021} and \citet{zhang25}. The duration of the simulation $t_\mathrm{f}$ is consistent with the estimated age of the bar of $\approx8$\,Gyr \citep{sanders2024}. We consider a range of present-day pattern speeds $\Omega_\mathrm{b,0}$. As the bar slows, its amplitude is held constant while its length scales according to $S(t)=\Omega_\mathrm{b,0}/\Omegab(t)$, where $S=1$ corresponds to the \citet{hunter2024} model. This ensures that the present-day potential matches this model for all $\Omega_\mathrm{b,0}$.

We perform two sets of simulations in this potential: a) integrating a set of particles from a GSE-like distribution function forwards from $t=0$ to $t=t_\mathrm{f}$; and b) integrating the sample of \ocen phase space positions back in time from the present-day $t=t_\mathrm{f}$ to $t=0$. These will determine whether the bar is capable of transporting particles from the GSE to the low-energy retrograde region of phase space occupied by \ocen.
\begin{figure*}
  \centering
  \includegraphics[width=\textwidth]{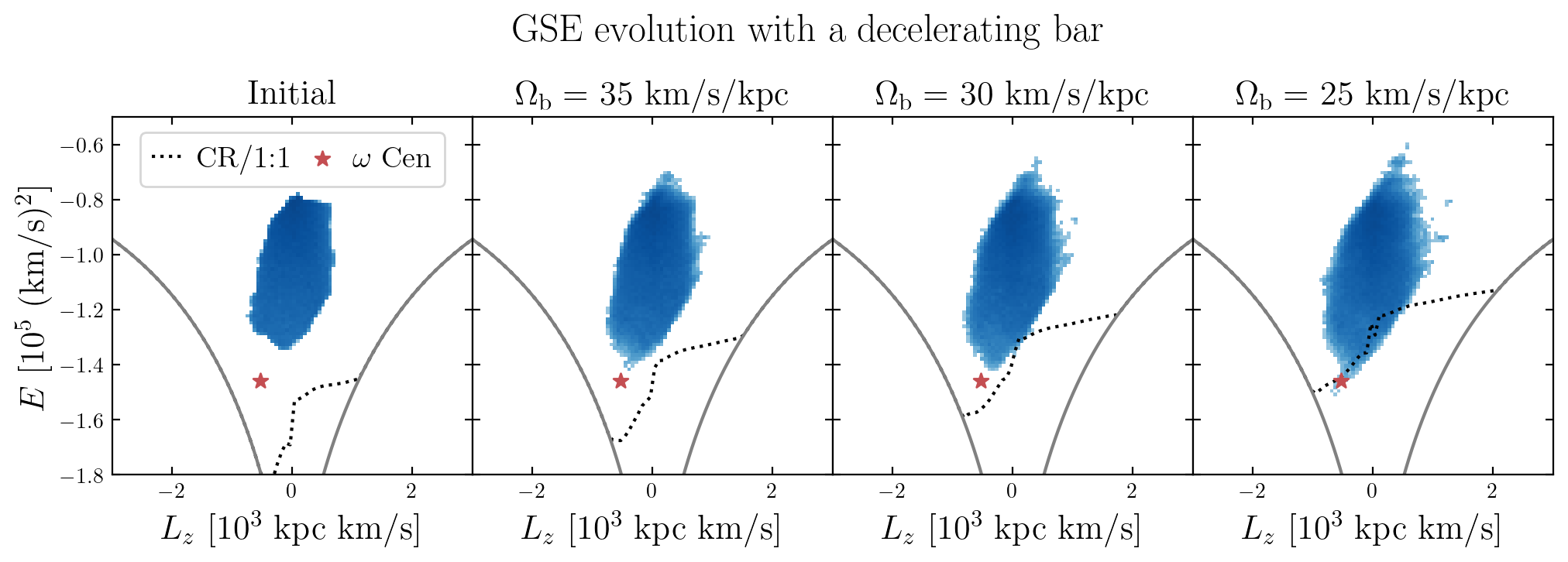}
  \caption{Evolution of a GSE-like population in $(L_z,E)$ space subject to a decelerating bar. The position of \ocen is marked with a star symbol, and the prograde CR/retrograde 1:1 resonance is marked with a dotted line. Each panel shows a different snapshot, with $\Omegab$ decreasing from left to right. GSE particles only reach the location of \ocen when $\Omegab\approx25$~km\,s$^{-1}$\,kpc$^{-1}$, at which point \ocen lies close to the resonance.}
   \label{fig:gse_evolution}
\end{figure*}
\begin{figure*}
  \centering
  \includegraphics[width=\textwidth]{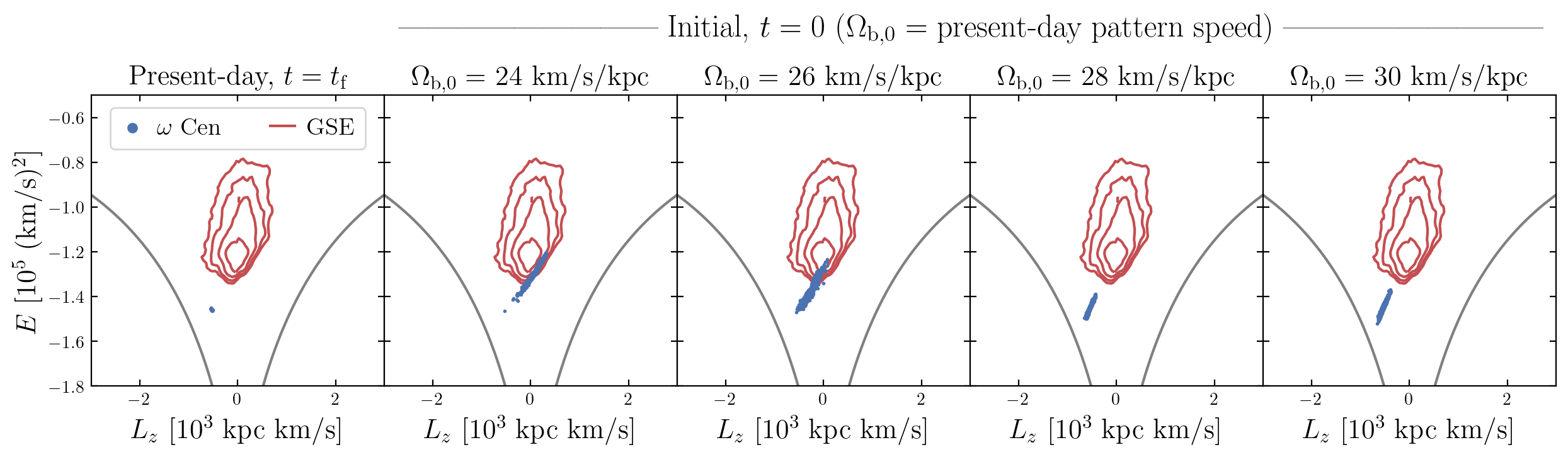}
  \caption{Location in $(L_z,E)$ space of \ocen in a selection of simulations. \textbf{Left-hand panel:} the present-day snapshot, showing samples of \ocen's location from observational uncertainties. The contours show the GSE debris, taken from \citet{belokurov_chevrons}. \textbf{Other panels:} initial snapshots of the simulations, after the orbits have been integrated backwards from the samples in a barred potential. Each panel shows a simulation with a different present-day pattern speed $\Omega_\mathrm{b,0}$ (indicated in each heading). There is significant overlap between the \ocen probability distribution and the GSE contours when $\Omega_\mathrm{b,0}\lesssim26$~km\,s$^{-1}$\,kpc$^{-1}$.}
   \label{fig:oCen_evolution}
\end{figure*}
\begin{figure}
  \centering
  \includegraphics[width=\columnwidth]{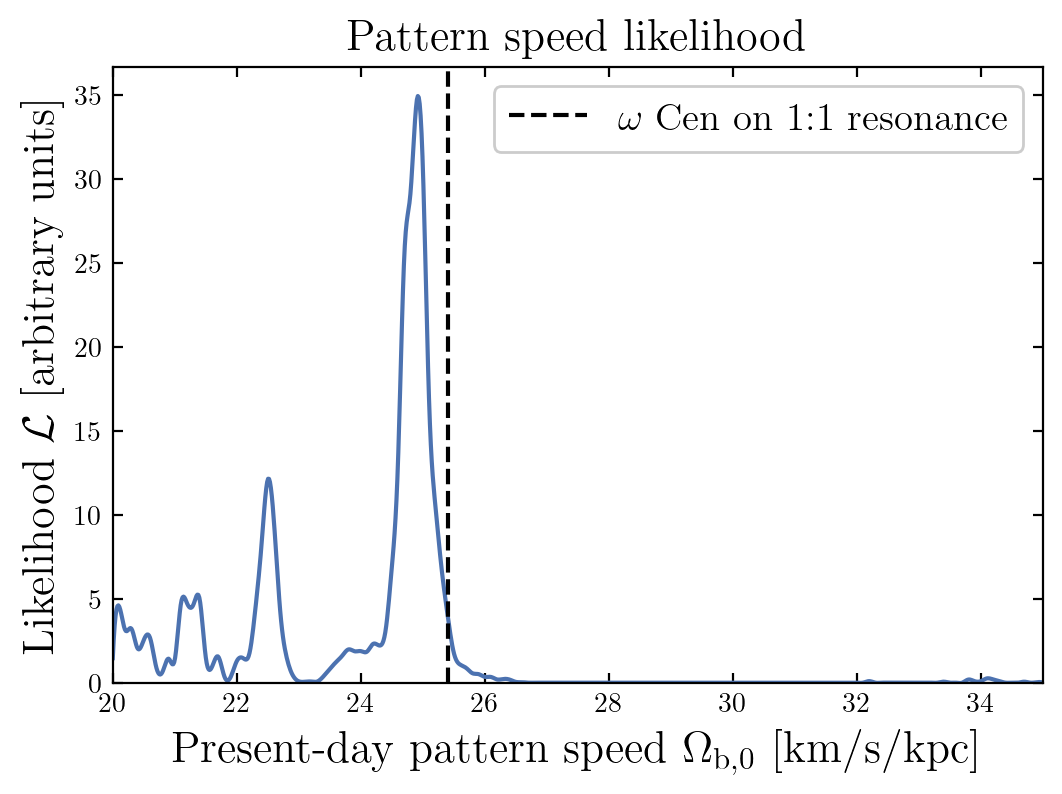}
  \caption{Likelihood of \ocen's phase space position as a function of the present-day pattern speed of a decelerating bar. The vertical dashed line is the pattern speed at which \ocen's frequencies satisfy $\Omega_\phi-\Omegab+\Omega_r=0$, so that it sits on the 1:1 resonance. There is only significant likelihood when $\Omega_\mathrm{b,0}$ is lower than this value, meaning that the resonance has passed the location of \ocen in the past.}
   \label{fig:likelihood}
\end{figure}

\noindent\textbf{GSE evolution.} We represent the GSE with a distribution function (DF) based on the combination Osipkov-Merritt model described by \citet{lane2025}. This consists of a superposition of two Osipkov-Merritt DFs \citep[see][]{binney_tremaine} whose anisotropies $\beta$ vary between 0 at the Galactic centre and 1 at $r\to\infty$. \citet{lane2025} fitted this DF to GSE stars and demonstrated that it provides a good fit across a wide range of radii. However, we found that its stars extend to lower energy than predicted for the GSE by \citet{belokurov_chevrons}. We therefore truncate the DF in $(L_z,E)$ space by setting it to zero outside the GSE contours shown in Fig.~\ref{fig:metallicity}. While this may result in an underestimated density at high energies, it ensures that the low-energy extent of the GSE in action space is as realistic as possible. We denote this DF in terms of the actions $\boldsymbol{J}$ as $f_\mathrm{gse}(\boldsymbol{J})$.

We draw a sample of $\approx10^5$ stars from $f_\mathrm{gse}$ and integrate their orbits from $t=0$ to $t=t_\mathrm{f}$ in the decelerating barred potential. We use a final pattern speed of $\Omega_\mathrm{b,0}=24$~km\,s$^{-1}$\,kpc$^{-1}$, which gives an initial pattern speed of $\approx45$~km\,s$^{-1}$\,kpc$^{-1}$. We show the distribution of the particles in $(L_z,E)$ space at different snapshots in Fig.~\ref{fig:gse_evolution}. For comparison we also mark the location of \ocen with a star symbol, and the prograde CR/retrograde 1:1 resonance with a dotted line. The left-hand panel shows the initial snapshot, where particles are all contained within the GSE contours from \citet{belokurov_chevrons}. The other three panels show snapshots with pattern speeds $\Omegab=\{35,30,25\}$ km\,s$^{-1}$\,kpc$^{-1}$.

As the bar decelerates, the distribution of particles in $(L_z, E)$ space spreads out. This dispersal occurs along lines of gradient $\Omegab$, as discussed by \citet{dillamore2025b}. However, particles only reach the location of \ocen when pattern speed has decreased to $\Omegab\approx25$~km\,s$^{-1}$\,kpc$^{-1}$. This approximately corresponds to the time at which the retrograde 1:1 resonance reaches the lower edge of the GSE distribution and \ocen itself. As explained by \citet{tomlinson2026}, this specific resonance is capable of scattering particles along its length onto significantly more retrograde orbits. This process allows GSE stars to reach the phase space location of \ocen. For this to have occurred, the retrograde 1:1 resonance must have been aligned with the location of \ocen at some point in the bar's past evolution. As demonstrated by Fig.~\ref{fig:gse_evolution}, this requires that the pattern speed has been as slow as $\Omegab<30$~km\,s$^{-1}$\,kpc$^{-1}$. This is considerably lower than most estimates of the current value of $\Omegab$ \citep[typically $30-45$~km\,s$^{-1}$\,kpc$^{-1}$; see figs. 10 and 2 in][respectively]{hunt2025,eugene2026}, although it is consistent with \citet{horta2024}.

In summary, Fig.~\ref{fig:gse_evolution} provides qualitative evidence that the location of \ocen is consistent with an origin in the GSE, provided that the pattern speed has been sufficiently slow to cause migration. In the next subsection we quantitatively estimate the pattern speed required for GSE stars to diffuse to the location of \ocen.

\noindent\textbf{\ocen evolution.} We now run a complementary set of simulations, where we instead integrate the orbit of \ocen back in time. We start from the samples of its observed position $(\boldsymbol{x}_i,\boldsymbol{v}_i)$ described above. We integrate backwards in barred potentials with different present-day pattern speeds $\Omega_\mathrm{b,0}$. Each simulation has the same deceleration rate $\eta=0.003$ and integration time. In Fig.~\ref{fig:oCen_evolution} we show the results of a selection of these simulations in $(L_z,E)$ space. The left-hand panel shows the observed (present-day) samples of \ocen's location, with the GSE contours from \citet{belokurov_chevrons} for comparison. The other panels show the initial snapshot at $t=0$ (after integrating the orbits back in time). The title of each panel indicates the present-day pattern speed $\Omega_\mathrm{b,0}$. 

Fig.~\ref{fig:oCen_evolution} clearly shows that the probability distribution of \ocen's initial phase space location depends strongly on the pattern speed. A strong overlap with the GSE contours requires a value of $\Omega_\mathrm{b,0}\lesssim26$~km\,s$^{-1}$\,kpc$^{-1}$. This is consistent with the expectations from Fig.~\ref{fig:gse_evolution}, and is again due to the location of the retrograde 1:1 resonance. If the pattern speed is any higher, the energy and angular momentum of \ocen remain approximately constant throughout the simulation. In this case bar-induced migration alone is not sufficient to transport \ocen away from the GSE debris.

\noindent\textbf{Pattern speed constraint.} We can use these results to quantitatively constrain the value of $\Omegab$ consistent with the migration of \ocen from the GSE debris. Let $f_\mathrm{gse}(\boldsymbol{J})$ be the distribution function (DF) of the GSE debris before bar formation, assumed to be in equilibrium. The bar causes this to evolve to the present-day DF $f_0(\boldsymbol{x},\boldsymbol{v}|\Omega_\mathrm{b,0})$. This notation indicates that the final DF depends on the pattern speed; in this study we vary the present-day value $\Omega_\mathrm{b,0}$ while keeping the deceleration rate and bar's age constant.

This can be used to define a likelihood for the phase space location of \ocen, on the assumption that it originated in GSE. We assume no prior knowledge of its initial location within the GSE debris, so the probability distribution of its phase space position before bar formation is proportional to $f_\mathrm{gse}$. The probability of its current position is therefore described by $f_0$. This gives a likelihood for the location of \ocen,
\begin{align}
    \mathcal{L}\propto\frac{1}{N}\sum_{i=1}^Nf_0(\boldsymbol{x}_i,\boldsymbol{v}_i|\Omega_\mathrm{b,0}),
\end{align}
where the sum is over all samples of the observed phase space location of \ocen.

Due to the collisionless Boltzmann equation $\mathrm{d}f/\mathrm{d}t=0$, the initial and final DFs $f_\mathrm{gse}$ and $f_0$ are equal when evaluated on the same orbit. We can therefore compute the likelihood by taking the samples $(\boldsymbol{x}_i,\boldsymbol{v}_i)$; integrating their orbits \textit{backwards} in time from $t=t_\mathrm{f}$ to $t=0$; calculating their actions $\boldsymbol{J}_i$ at $t=0$ in the axisymmetrised potential; and computing the DF $f_\mathrm{gse}(\boldsymbol{J}_i)$. This is computationally much cheaper than a forward integration from samples of $f_\mathrm{gse}$, because only the small region of phase space occupied by \ocen needs to be sampled. This gives a new expression for the likelihood,
\begin{align}
    \mathcal{L}\propto\frac{1}{N}\sum_{i=1}^Nf_\mathrm{gse}(\boldsymbol{J}_i(\boldsymbol{x}_i,\boldsymbol{v}_i|\Omega_\mathrm{b,0})),
\end{align}
where we indicate that $\boldsymbol{J}_i$ are functions of the present-day coordinates $(\boldsymbol{x}_i,\boldsymbol{v}_i)$, with the pattern speed as the single model parameter.

We plot the likelihood $\mathcal{L}$ as a function of $\Omega_\mathrm{b,0}$ in Fig.~\ref{fig:likelihood}. Assuming a uniform prior in $\Omega_\mathrm{b,0}$, this is equivalent to a posterior probability distribution for the pattern speed within our model assumptions. The vertical dashed line marks the pattern speed at which \ocen sits exactly on the retrograde 1:1 resonance, such that its frequencies satisfy $\Omega_\phi-\Omegab+\Omega_r=0$. Fig.~\ref{fig:likelihood} confirms that \ocen can likely only be traced back to the GSE debris if the pattern speed is around or lower than this critical value, or $\Omega_\mathrm{b,0}\lesssim26$~km\,s$^{-1}$\,kpc$^{-1}$. If the pattern speed has never been this slow, the 1:1 resonance cannot have caused sufficient migration of \ocen. The likelihood profile has many peaks and troughs at these low pattern speeds; these are likely dependent on details of the simulation such as bar deceleration rate, which we keep fixed. Thus details of the profile in Fig.~\ref{fig:likelihood} should not be over-interpreted.

We also note that we have only considered a decelerating bar in this model. In this case $\Omega_\mathrm{b,0}$ is the slowest pattern speed throughout the bar's evolution. While this is the most likely scenario \citep[e.g.][]{tremaine1984,athanassoula2003,chiba2021}, it is possible that the pattern speed undergoes fluctuations including brief periods of acceleration \citep[e.g.][]{hilmi2020}. We should thus interpret our upper bound as a constraint on the \textit{minimum} pattern speed of the bar throughout its history. We therefore argue that the \ocen migration scenario requires that the minimum pattern speed of the bar satisfies $\Omega_\mathrm{b,min}\lesssim26$~km\,s$^{-1}$\,kpc$^{-1}$.

\section{Discussion and conclusions}\label{section:conclusions}

\noindent\textbf{Comparison with previous pattern speed constraints.} Most recent estimates of the pattern speed are in the range $\Omegab=30-45$~km\,s$^{-1}$\,kpc$^{-1}$ \citep{Po17,Sa19,bovy2019,binney2020,chiba2021_treering,Cl22,li2022,leung2023,zhang24,dillamore2025}. Our upper bound of $\approx26$~km\,s$^{-1}$\,kpc$^{-1}$ inferred from \ocen's migration is therefore much slower than the general consensus. However, some works have found lower values; \citet{horta2024} estimate $\Omegab=24\pm3$~km\,s$^{-1}$\,kpc$^{-1}$, in good agreement with our result. Hence there remains considerable uncertainty about the pattern speed. However, if values below 26~km\,s$^{-1}$\,kpc$^{-1}$ can be definitively ruled out by future studies, this would help to falsify the bar-induced migration scenario.

\noindent\textbf{Dynamical friction.} We did not include dynamical friction on \ocen in our simulations. \citet{moreno2022} estimated the rates of change of $L_z$ and $E$ due to dynamical friction for a set of globular clusters in the Milky Way. For \ocen they found $\dot{L}_z\approx9.55$~kpc\,km\,s$^{-1}$\,Gyr$^{-1}$ and $\dot{E}\approx-6.27\times10^{2}$~km$^2$\,s$^{-2}$\,Gyr$^{-1}$. Over $\sim10$~Gyr these rates are too slow to move the cluster from the GSE debris to its current energy. It is possible that \ocen has lost mass over time, so the energy loss due to dynamical friction could be greater than this estimate. However, \citet{moreno2022} also found that dynamical friction causes $|L_z|$ to decrease over time for \ocen, so bar-induced migration must still be invoked to explain its retrograde orbit. Including dynamical friction in our model would likely decrease the upper bound on the pattern speed, since the 1:1 resonance would have to be at an even higher energy. Our estimate of the bound without dynamical friction is therefore conservative. Nevertheless, incorporating dynamical friction into the model may be worthwhile in future work.

\noindent\textbf{Summary.} We have investigated the scenario in which the globular cluster \ocen originated in the GSE merger event (e.g. as a nuclear star cluster), and subsequently migrated to its current orbit due to resonance with the Galactic bar \cite{laporteorkney2026}.

Extending the chemical analysis of \citet{laporteorkney2026}, we find that the proposed metal-poor corridor between the GSE debris and \ocen is not a chemically simple continuation of the main low-$\alpha$ GSE sequence. The low-$\alpha$ component is clearest near the GSE cloud and only tentatively reaches \ocen in the bridge slice that contains the cluster; elsewhere the bridge region is more $\alpha$-enhanced than the GSE core. The chemical evidence therefore remains suggestive rather than conclusive: it is compatible with a GSE origin for \ocen, but does not require one.

The retrograde 1:1 resonance is capable of causing migration of stars from the GSE debris to the current orbit of \ocen. This requires that the minimum pattern speed of the bar throughout its evolution was $\Omega_\mathrm{b,min}\lesssim26$~km\,s$^{-1}$\,kpc$^{-1}$. This is considerably slower than most estimates of the current value (typically $\Omegab=30-45$~km\,s$^{-1}$\,kpc$^{-1}$), though consistent with \citet{horta2024}.

In summary, the GSE/\ocen migration scenario is plausible, but would require a re-evaluation of the general consensus on the bar's pattern speed.

\section*{Acknowledgements}
We thank the anonymous referee for a helpful report. AMD acknowledges support from the Royal Society (URF\textbackslash R1\textbackslash191555; URF\textbackslash R\textbackslash 241030). HZ thanks the Science and Technology Facilities Council (STFC) for a PhD studentship (grant number 2888170). VB is grateful for the support from the Leverhulme Trust Research Project Grant RPG-2021-205 `The Faint Universe Made Visible with Machine Learning'.

\section*{Data Availability}
This study uses publicly available \textit{Gaia} data. The code used in this project can be found at \url{https://github.com/adllmr/oCen_bar}.



\bibliographystyle{mnras}
\bibliography{refs} 




\appendix


\bsp	
\label{lastpage}
\end{document}